\documentclass[aps,prb,twocolumn,showpacs,preprintnumbers,prb,superscriptaddress]{revtex4-1}
\usepackage{graphicx}  
\usepackage{dcolumn}   
\usepackage{bm}        
\usepackage{amssymb}   
\usepackage{framed}
\usepackage{amsmath}
\usepackage{ulem}    
\usepackage{hhline}
\usepackage{subfigure,amsmath,verbatim,moreverb}
\usepackage{tabularx}
\usepackage{lipsum}
\usepackage{longtable}
\usepackage{booktabs}
\usepackage{threeparttable}
\usepackage{booktabs,amssymb,siunitx}
\usepackage{latexsym}
\usepackage{dcolumn}
\usepackage{amsmath}
\usepackage{epsf}
\usepackage{float}
\usepackage{hyperref}
\usepackage{xcolor}
\usepackage{etoolbox}
\AtBeginEnvironment{align}{\setcounter{subeqn}{0}}
\newcounter{subeqn} %

\begin{document}
\title{Improved electronic structure prediction of chalcopyrite semiconductors from a semilocal density functional based on Pauli 
kinetic energy enhancement factor}
\author{Arghya Ghosh}
\email{ph17resch11006@iith.ac.in}
\affiliation{Department of Physics, Indian Institute of Technology, Hyderabad, India}
\author{Subrata Jana}
\email{subrata.jana@niser.ac.in, subrata.niser@gmail.com}
\affiliation{School of Physical Sciences, National Institute of Science Education and Research, HBNI, 
Bhubaneswar 752050, India}
\author{Manish K Niranjan}
\email{manish@phy.iith.ac.in}
\affiliation{Department of Physics, Indian Institute of Technology, Hyderabad, India}
\author{Sushant Kumar Behera}
\affiliation{School of Physical Sciences, National Institute of Science Education and Research, HBNI, 
Bhubaneswar 752050, India}
\author{Lucian A. Constantin}
\affiliation{Istituto di Nanoscienze, Consiglio Nazionale delle Ricerche CNR-NANO, 41125 Modena, Italy}
\author{Prasanjit Samal}
\affiliation{School of Physical Sciences, National Institute of Science Education and Research, HBNI, 
Bhubaneswar 752050, India}

\date{\today}

\begin{abstract}

The correct treatment of $d$ electrons is of prime importance in order to predict
the electronic properties of the prototype chalcopyrite semiconductors. The effect 
of $d$ states is linked with the anion displacement parameter $u$, which in turn 
influences the bandgap of these systems. Semilocal exchange-correlation functionals which yield good 
structural properties of semiconductors and insulators often fail to predict 
reasonable $u$ because of the underestimation of the bandgaps arising from the 
strong interplay between $d$ electrons. In the present study, we show that the 
meta-generalized gradient approximation (meta-GGA) obtained from the cuspless 
hydrogen density (MGGAC) [Phys. Rev. B 100, 155140 (2019)] performs in an improved 
manner in apprehending the key features of the electronic properties of chalcopyrites, 
and its bandgaps are comparative to that obtained using state-of-art hybrid 
methods. Moreover, the present assessment also shows the importance of the Pauli 
kinetic energy enhancement factor, $\alpha=(\tau-\tau^W)/\tau^{unif}$ in describing
the $d$ electrons in chalcopyrites. The present study strongly suggests that
the MGGAC functional within semilocal approximations can be a better and preferred 
choice to study the chalcopyrites and other solid-state systems due to its superior 
performance and significantly low computational cost.

\end{abstract}

\maketitle

\section{Introduction}

In the last two decades, the $ABX_2$ (A=Cu,Ag,Be,Cd,Mg,Zn; B=Ga,Ge,In; X=S/Se/Te) chalcopyrite semiconductors 
based on the zinc-blend structure have emerged as highly promising advanced solar cells materials that allow 
the tweaking of macroscopic physical properties as function of microscopic chemical and structural degrees of 
freedoms~\cite{green2019solar,jackson2011new,walsh2012kesterite,feng2011three,rife1977optical,alonso2001optical}. 
In particular, Cu-based ternary and quaternary semiconductors and their alloys are now widely exploited as 
light absorbers in thin film solar cells~\cite{birkmire1997poly,repins2008progress,jackson2011new,green2012solar,
rashkeev2001second} and various non liner optical devices~\cite{spiess1974nuclear,parkes1073crystal,dittmer2000electron}. 
Unfortunately, the theoretical predictions of the structural and electronic properties of Cu-based multinary 
semiconductors remain far from being accurate because of the involvement of semicore $d$ or $f$ electrons, 
which play a decisive role in the determination of the properties of these materials~\cite{vidal2010strong,siebentritt2010electronic}.

The straightforward applications of the Kohn-Sham (KS) density functional theory 
(DFT)~\cite{kohn1965self} within the framework of the local density approximation (LDA)~\cite{perdew1992accurate} or generalized gradient 
approximation (GGA)~\cite{perdew1996generalized} often result in wrong predictions of bandgaps and band 
ordering as these suffer from the delocalization error~\cite{cohen2008insights}. Thus, the insufficiency and 
lack of appropriate theoretical approaches hinder the proper understanding of the inherent electronic properties 
of these systems. Although, some plausible attempts have been made to correlate bandgap with the 
physical parameters in chalcopyrites such as CuInSe$_2$, CuInS$_2$ in various literature~\cite{vidal2010strong,jiang2007electronic}, 
theoretical studies of such correlations and structural anomalies in relevant and important 
prototypes such as CuGaS$_2$, CuAlS$_2$, are still fragmentary.

Nevertheless, using hybrid functional schemes~\cite{heyd2003hybrid,heyd2004efficient,krukau2006influence,
paier2006screened,jana2018efficient,jana2019screened,jana2020screened,jana2020improved}, improved bandgaps of chalcopyrites
and other insulators have been reported. These schemes combine Hartree-Fock exchange with LDA/GGA 
correlation. Further, the semilocal exchange-correlation (XC) effect of strongly localized $d$-orbitals are 
treated at the mean-field level~\cite{anisimov1991density,Anisimov_1997,anisimov1991band,zhang2011comparative}. Although the hybrid density 
functionals have been applied and studied for a variety of systems, these schemes are often 
computationally expensive and prohibitive. On the other hand, the strong electron correlations and self-interaction 
related problems within semilocal approximation can be partially mitigated by the addition of separate Hubbard $U$ for 
the $d$ or/and $f$ localized electrons. This scheme may be expected to provide a qualitative assessment of the influence of the electronic 
correlations on the physical properties of a system~\cite{anisimov1991density,Anisimov_1997}. The DFT+U method attempts to fix hybridization 
problems due to overly-delocalized orbitals and thereby has been recognized as a potent tactic to treat the strongly localized 
$d$ electrons in chalcopyrites~\cite{anisimov1991band,zhang2011comparative}.

In addition to hybrid and DFT+U schemes, the quasi-particle correction method within the many-body 
GW approximation is even more well-founded among different levels of theory 
which renders a relatively more accurate estimation of bandgaps~\cite{anisimov1991density,Anisimov_1997}. 
However, the calculations using the GW scheme are usually more computationally expensive than those using
hybrid functionals.

Over the years, the search for efficient semilocal approaches which can predict the accurate electronic structure of systems, has become an important and fascinating research topic. The progress in 
semilocal density functionals in recent years shows that the accuracy of the material properties 
can be achieved from the advanced XC methods by satisfying more exact quantum mechanical constraints.
The advent of these accurate semilocal density functionals~\cite{sun2015strongly,
tao2016accurate,patra2019relevance,patra2019efficient,patra2020way,furness2020accurate} has also 
led to wide applicability of DFT in the condensed matter physics. 
In particular, the meta-GGA functionals have been found to be quite successful 
in overcoming the several challenges associated with the lower rung of functionals such as GGA, 
LDA etc.~\cite{jana2019improving,jana2018assessment,
jana2018assessing,zhang2020symmetry,zhang2020competing,nokelainen2020abinitio,patra2021correct}

Though the chalcopyrite systems have been studied using different functionals, the assessment, and performance 
of the advanced meta-GGA methods for these systems have not been reported to the best of our knowledge. Therefore, 
in this article, we investigate the performance of some advanced meta-GGA functionals such as 
1) meta-GGA functional obtained from the 
cuspless hydrogen density (MGGAC)~\cite{patra2019relevance},
2) strongly constrained 
and appropriately normed (SCAN) semilocal functional~\cite{sun2015strongly} , and 3) Tao-Mo (TM)~\cite{tao2016accurate} semilocal functional.
We assess the performance of these functionals in the study of structural and electronic properties, 
bandgaps, and enthalpy formation energies of chalcopyrite systems. It may be noted that both SCAN and MGGAC functionals have been found 
to be quite successful in overcoming the several drawbacks of the GGA functional such as severe underestimation of the bandgaps of the 
bulk and layered solids~\cite{patra2020electronic}. In particular, our study shows that
the MGGAC functional is quite advantageous to study the chalcopyrite semiconductors as it is computationally inexpensive and exhibits 
good performance. The aforementioned meta-GGA schemes span the third rung of the Jacob ladder of the XC approximation~\cite{perdew2001jacob} owing to 
important features associated with them.

This paper is organized as follows. The methodologies used in this work are discussed in section II. The structural properties, 
electronic properties, and enthalpy formation energies are discussed in sections II, III, and IV respectively. The results and 
the functional performances are further analyzed in section VI. Finally, the conclusions are presented in section VI.

\section{Computational Methods}

The density functional calculations are performed using the plane-wave formalism  
as implemented in Vienna Ab-initio Simulation Package (VASP) code~\cite{PhysRevB.47.558,PhysRevB.54.11169,PhysRevB.59.1758,KRESSE199615}. 
The core-valence electron interaction is approximated using the projected augmented wave (PAW) method~\cite{paw}. 
A kinetic energy cutoff of 600 eV is used to expand the Kohn-Sham (KS) single-particle orbitals on a plane-wave basis.
The Brillouin Zones (BZ) are sampled using 
Monkhorst-Pack $9\times 9\times 9$ {\bf{k}}-points grid for atomic relaxations and 17$\times$17$\times$17 grid  
 for the density of states (DOS) calculations. 
The electronic energies are allowed to converge up to 
10$^{-6}$ eV (or less) to achieve self-consistency in the calculations.  
The atomic relaxations are performed till the
Hellmann-Feynman forces on atoms are 
reduced to less than $0.01$ eV/\AA. 

The exchange-correlation (XC) contributions are included using semilocal XC functional schemes such as PBE~\cite{perdew1996generalized}, 
SCAN~\cite{sun2015strongly}, MGGAC~\cite{patra2019relevance}, and TM~\cite{tao2016accurate}. Unlike GGAs, the meta-GGA functionals SCAN, 
MGGAC, and TM are implemented in generalized KS (gKS)~\cite{perdew2017understanding,yang2016more} scheme because of their dependence on 
KS orbitals. Usually, the meta-GGAs are quite reliable as compared to GGAs as they can recognize and describe covalent, metallic, and 
non-covalent interactions ~\cite{sun2013density,della2016kinetic}.

In general, the meta-GGA XC functional which depends on the density ($\rho$), gradient of density ($\nabla\rho$), and 
KS kinetic energy density ($\tau({\bf{r}})=\frac{1}{2}\sum_{i}|\nabla\psi_{i}|^2$) 
can be presented as:
\begin{equation}
 E_{xc}[\rho_{\uparrow},\rho_{\downarrow}]=\int~d^3r~\rho({\bf{r}})\epsilon_{x}^{LDA}F_{xc}(\rho_{\uparrow},\rho_{\downarrow},
 \nabla\rho_{\uparrow},\nabla\rho_{\downarrow},\tau_{\uparrow},\tau_{\downarrow})~,
 \label{introeq1}
\end{equation}
where $\epsilon_{x}^{LDA}$ is the exchange energy density in uniform electron gas approximation and $F_{xc}$ is the XC enhancement
factor. In case of construction of SCAN~\cite{sun2015strongly} and MGGAC~\cite{patra2019relevance} functionals, one of the main 
ingredients in $F_x$ is the Pauli kinetic enhancement factor $\alpha=(\tau-\tau^{W})/\tau^{unif}$, where $\tau^{W}=|\nabla\rho|/[8\rho]$ is the 
von Weizs\"{a}cker kinetic energy 
density, $\tau^{unif}=3/10(3\pi^2)^{2/3}\rho^{5/3}$ is the Thomas-Fermi kinetic energy density, and $k_F=(3\pi^2\rho)^{1/3}$ is the Fermi wave vector. It may be noted that the exchange of the SCAN functional 
uses both the $\alpha$ and $p=s^2$, while the exchange in MGGAC uses only $\alpha$. This makes MGGAC $F_{xc}$ in exchange-only or 
high-density limit ($r_s\to 0$) independent of $s$. Importantly, both SCAN and MGGAC respect the tightened bound of the 
exchange ($F_x \leq 1.174$~\cite{perdew2014gedanken,sun2015strongly}) and possess negative slope with respect to $\alpha$ 
($\partial F_x/\partial\alpha < 0$), important for the bandgap improvement in semilocal level~\cite{aschebrock2019ultranonlocality,
patra2020electronic,patra2019relevance}.

Another reliable meta-GGA semilocal functional was proposed by Tao-Mo (TM)~\cite{tao2016accurate}
which has been quite successful at providing reasonably accurate estimates of various quantum chemical and solid-state 
properties~\cite{mo2017assessment,mo2017performance,tao2016accurate,jana2019improving,jana2020insights}. The TM and its 
revised version revTM\cite{jana2019improving} functionals, use both the meta-GGA ingredients i.e., $z=\tau^W/\tau$ and $\alpha$. 
However, these functionals satisfy looser bound of exchange~\cite{lewin2015improved,doi:10.1063/1.4904448,constantin2015gradient} and 
suffer from the order-of-limit problem~\cite{patra2020way}.

\begin{table}
\scriptsize
\begin{center}
\caption{\label{tab-all-0} I-III-VI2 and I-IV-V2 chalcopyrite semiconductor systems considered in this paper.}
\begin{ruledtabular}
\begin{tabular}{l||cccccccccccccccccccccccc}
Group&Systems\\
\hline
         &CuInS$_2$, CuInSe$_2$, CuGaS$_2$, CuGaSe$_2$, CuAlS$_2$, CuAlSe$_2$,\\
I-III-VI2        &CuAlTe$_2$, CuGaTe$_2$, CuInTe$_2$, AgAlS$_2$, AgAlSe$_2$, AgAlTe$_2$,\\
        &AgGaS$_2$, AgGaSe$_2$, AgGaTe$_2$, AgInS$_2$, AgInSe$_2$, AgInTe$_2$\\
        \hline\hline
                 &BeSiP$_2$, BeSiAs$_2$, BeGeP$_2$, BeGeAs$_2$, BeSnP$_2$, BeSnAs$_2$\\
II-IV-V2          &MgSiP$_2$, MgSiAs$_2$, MgGeP$_2$, MgGeAs$_2$, MgSnP$_2$, MgSnAs$_2$\\
                 &ZnSiP$_2$, ZnSiAs$_2$, ZnGeP$_2$, ZnGeAs$_2$, ZnSnP$_2$, ZnSnAs$_2$\\
                 &CdSiP$_2$, CdSiAs$_2$, CdGeP$_2$, CdGeAs$_2$, CdSnP$_2$, CdSnAs$_2$
\end{tabular}
\end{ruledtabular}
\end{center}
\end{table}

In addition to aforementioned functionals, we have also used the hybrid functional HSE06 scheme
propounded by Hyde, Scuseria, and Ernzerhof~\cite{heyd2003hybrid,heyd2004efficient, krukau2006influence,paier2006screened}. 
The HSE06 scheme is based on the inclusion of fixed amount of Hartree-Fock exchange. Briefly, the 
exchange-correlation energy of the HSE06 functional is given by~\cite{heyd2003hybrid,heyd2004efficient,krukau2006influence,
paier2006screened},

\begin{eqnarray}
E_{xc}^{HSE}\left(\alpha,\omega\right)&=&\alpha
E_x^{HF,SR}\left(\omega\right)+\left(1-\alpha\right)E_x^{PBE,SR}\left(\omega\right)\nonumber\\
&+&E_x^{PBE,LR}\left(\omega\right)+E_c^{PBE}~.
\end{eqnarray}

Here, the exchange-correlation energy $E_{xc}$ is range separated with the use of a screening potential. The parameter $\alpha$ describes the 
fraction of the Fock exchange while $\omega$ calibrates the range of the interaction. We have used the conventional $\alpha=0.25$
and $\omega=0.11$ bohr$^{-1}$ in computations with HSE06~\cite{krukau2006influence,paier2006screened}. It may be noted that other variants 
of hybrid density funtionals~\cite{jana2018efficient,jana2019screened,jana2020screened,jana2020improved} have also been proposed. 

In this work, we report the relative performance of HSE06, MGGAC, PBE, SCAN, and TM functionals by comparing the computed estimates of
the structural parameters and properties of the Cu based chalcopyrite systems. In case of other systems, the structural properties are 
calculated only using MGGAC, PBE, SCAN, and TM functionals and the results are reported in supplementary material~\cite{support}. 
Since, the main focus of the present work is the study of Cu based chalcopyrites, we do not relax the structures
of Ag, Be, Cd, Mg, and Zn based chalcopyrites using HSE06 scheme, which is quite expensive computationally. 
For these systems the structures are optimized and lattice constants are computed using PBE (GGA) scheme. 

In Table~\ref{tab-all-0}, we list the I-III-VI2 and II-IV-V2 group systems which are considered in present work.

\section{structural properties}

The chalcopyrite lattice structure adopts the  $D_{2d}^{12}$ (No.122) space group symmetry. The structure 
is a cation mutated super-structure of the cubic zinc-blende structure ($T^2_d$), in which each of the two 
cations $(A, B)$ are tetrahedraly coordinated by four anions ($X$), and each anion is coordinated by two $A$ 
and two type $B$ cations~\cite{jaffe1983electronic}. The chalcopyrite structure can be described accurately 
by three structural parameters: ($i$) lattice constant $a$, ($ii$) the tetragonal ratio $\eta=c/2a$ 
$\neq$ 1, where $c$  is the lattice constant in the $z$-direction, and ($iii$) the anion displacement parameter, $u$. 
In general, $A-X$ and $B-X$ bond lengths, denoted by $r_{A-X}$ and $r_{B-X}$, respectively, are not equal. 
This unequal anion-cation bond length results in a tetragonal distortion vis-a-vis an ideal zinc blende structure. 
The distortion parameter $u=0.25+(r^2_{A-X}-r^2_{B-X})/a^2$, which describes the re-positioning of the anions in the $x-y$ 
plane, sensitively influence the bulk electronic bandgap of these materials~\cite{jaffe1983electronic}.

The ternary ABC$_2$ [A = I; II = Cu, Ag; Be, Mg, Zn, Cd, B = III; IV = In, Ga, Al; Si, Ge, Sn, and C = VI; V = S, Se, Te; P, As]
chalcopyrite compounds of composition I$-$III$-$VI$_2$ or II$-$IV$-$V$_2$ 
are iso-electronic analogous of the II$-$VI or III$-$V binary semiconductors, respectively~\cite{shaposhnikov2012abinitio}. 
The crystal structure of these chalcopyrites is described by $D^{12}_{2d}$ ($I42d$) space group symmetry, 
and can be regarded as a super-lattice of the zinc-blende structure with small structural distortions~\cite{feng2011three}. 
The lattice constants ($a$ and $c$), distortion parameter ($u$), and tetragonal ratio ($\eta$)  of Cu based chalcopyrites 
computed using different XC schemes are listed in Table S1 of the supporting information~\cite{support}.
The trend seen in the results obtained from semilocal functionals may be explained from 
the perspective of computed electronic properties. 

In general, the computed lattice constants of solids obtained from 
PBE-like GGA XC functionals are
reasonably accurate. However, the PBE-GGA computed lattice constants are generally found slightly overestimated because of the lack of 
slowly varying density correction, which is taken care of in its solid-state version PBEsol~\cite{perdew2008restoring,terentjev2018dispersion}. 
Furthermore, the higher rung of functionals i.e., meta-GGAs are generally better than PBE-GGA 
functional for estimating the lattice constants. For instance, the SCAN meta-GGA
performs reasonablly well in predicting the lattice constants of different solids.

\begin{figure*}
      \includegraphics[width=11 cm, height=8.0 cm]{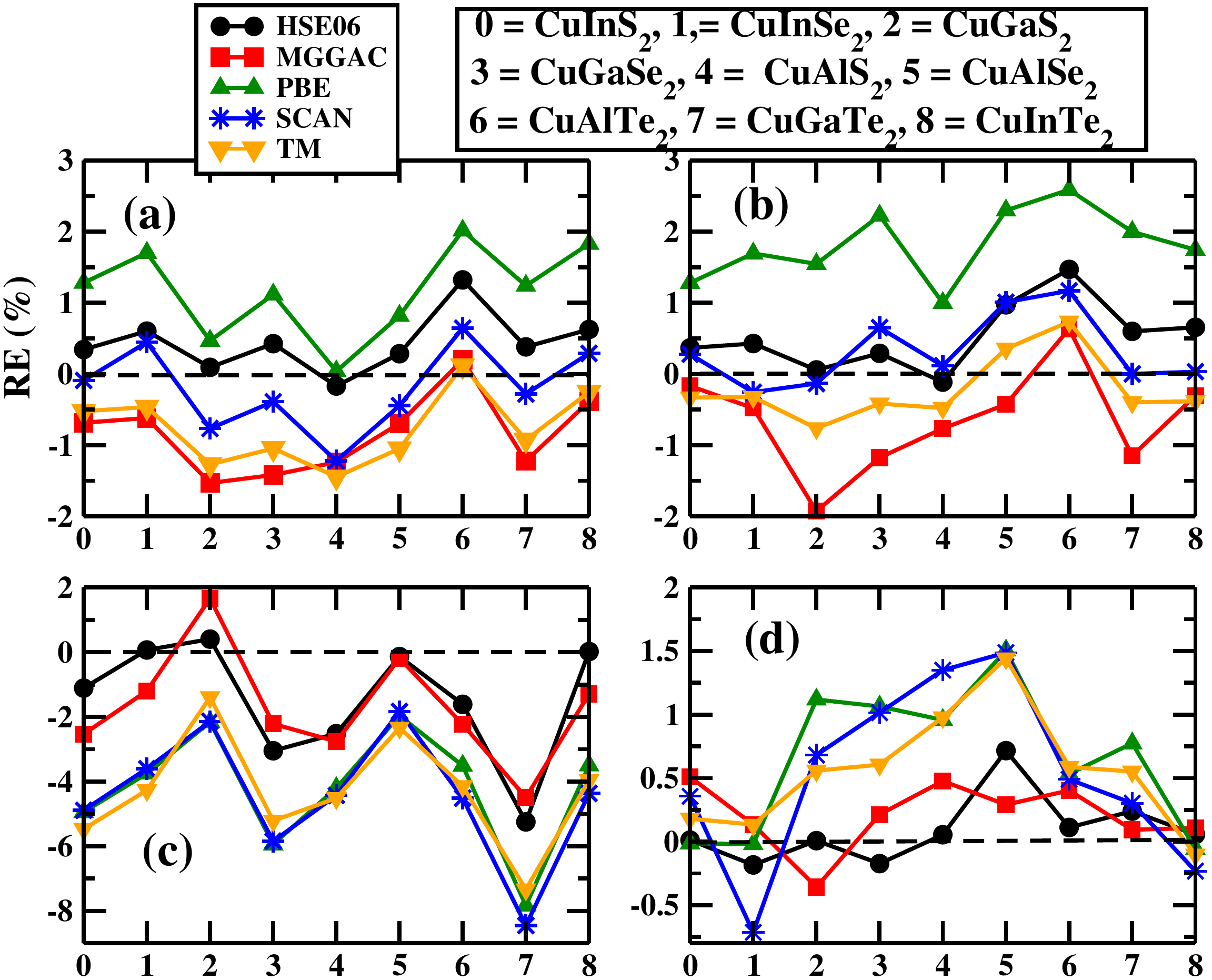}
    \caption{Relative error (\%) in (a) in-plane lattice constants, (b) lattice constants in $z$ direction, 
    (c) distortion parameter ($u$), and (d) tetragonal ratio ($\eta$) of the Cu based chalcopyrite systems 
    reported in Table S1 of the supporting information~\cite{support}. The levels along $x-$ axis is same as the solids presented 
    in Table S1 of the supporting information~\cite{support} and shown in the caption.}
  \label{fig-cu-systems}
\end{figure*}

In Fig~\ref{fig-cu-systems}, we show the relative error (in \%) in parameters $a$, $c$, $u$, and $\eta$ of the Cu based chalcopyrites as obtained from 
different methods. As can be seen, the HSE06 performs better than other methods. 
However, within semilocal methods, the SCAN performs better than MGGAC and PBE for $a$ and $c$, while MGGAC performs better than SCAN for $u$. 
Further, in most cases, the MGGAC underestimates $a$ and $c$ parameters as compared to experimental and SCAN values.  
It may also be noted that for the tetragonal ratio ($\eta$), the MGGAC performs in a more balanced way than SCAN due to
its tendency to underestimate both $a$ and $c$. Physically, the improvements in $u$ obtained using MGGAC as compared to those obtained using SCAN 
comes from bond lengths $r_{A/B-X}$, which are also significantly improved by much better description of the Cu $d$-states. 
It may also be seen in Fig~\ref{fig-cu-systems} that the $u$ parameters obtained from MGGAC are very close to those obtained using HSE06. This 
indicates that the performance of MGGAC is quite close to that of HSE06 in treating Cu-$d$ states. 
We will further discuss this point in more detail in section~\ref{elctronic-section}.

\section{electronic properties}
\label{elctronic-section}

The phenomenological treatment of $p$-$d$ interaction in chalcopyrite systems depends on how good are the various approximation of XC for 
the $d$-electron of Cu. For instance, the LDA (or GGA) based methods tend to delocalize $d$-electrons which results in underestimation of $p-d$ hybridization
~\cite{zhang2011comparative,vidal2010strong}.

Further, a significant shortcoming in standard DFT based methods is the presence of inherent self-interaction error (SIE)~\cite{vidal2010strong,
kim2016screened}. In the case of chalcopyrites, this shortcoming results in shallow $d$ states description for the cations, and in general,
the absence of derivative discontinuity in the KS bandgap which is defined as the difference between eigenvalues of 
the CBM and VBM. As a consequence, the theoretical bandgap is severely underestimated as compared to the experimental bandgap 
(the difference between the ionization potential and electron affinity). The absence of the LDA/GGA computed bandgaps in some of the 
chalcopyrites and other material systems has been attributed to large ratio of SIE in exchange functionals. Hybrid functionals 
eliminate some amount of SIE and improve the performance and estimates of bandgaps~\cite{vidal2010strong,kim2016screened}.
The performance of the meta-GGA functionals which are implemented in gKS scheme (such as SCAN and MGGAC), in providing estimates of the electronic properties of chalcopyrites has not been reported. We discuss these in next subsection.

To evaluate the relative performances of different functionals, we first discuss the band structures as obtained from HSE06, MGGAC, PBE, and SCAN
functionals. Following this, the density of states (DOS) and charge density will be analyzed and discussed. We have not included the performance
of the TM functional here as its performance is known to lie in between that of PBE and SCAN functionals. The bandgap values computed using TM functional 
are shown in Table S1 of the supporting information~\cite{support}. Overall, the performance of TM functional is found to qualitatively similar to that of the PBE 
functional.

In order to analyze the performance of different functionals, the band structures, charge density, and DOS of CuInS$_2$ are computed
using the HSE06, MGGAC, PBE, and SCAN functionals. In supporting information~\cite{support}, we also present and analyze the electronic properties of CuInSe$_2$. 
These two systems are particularly chosen since their electronic band structures, density of states (DOS), and charge densities are quite distinct 
from one another. Further, both these systems are known to exhibit strong interplay between structure and electronic properties~\cite{vidal2010strong}.


\subsection{ Band structures}

\begin{figure*}
      \includegraphics[width=10.0 cm, height=6 cm]{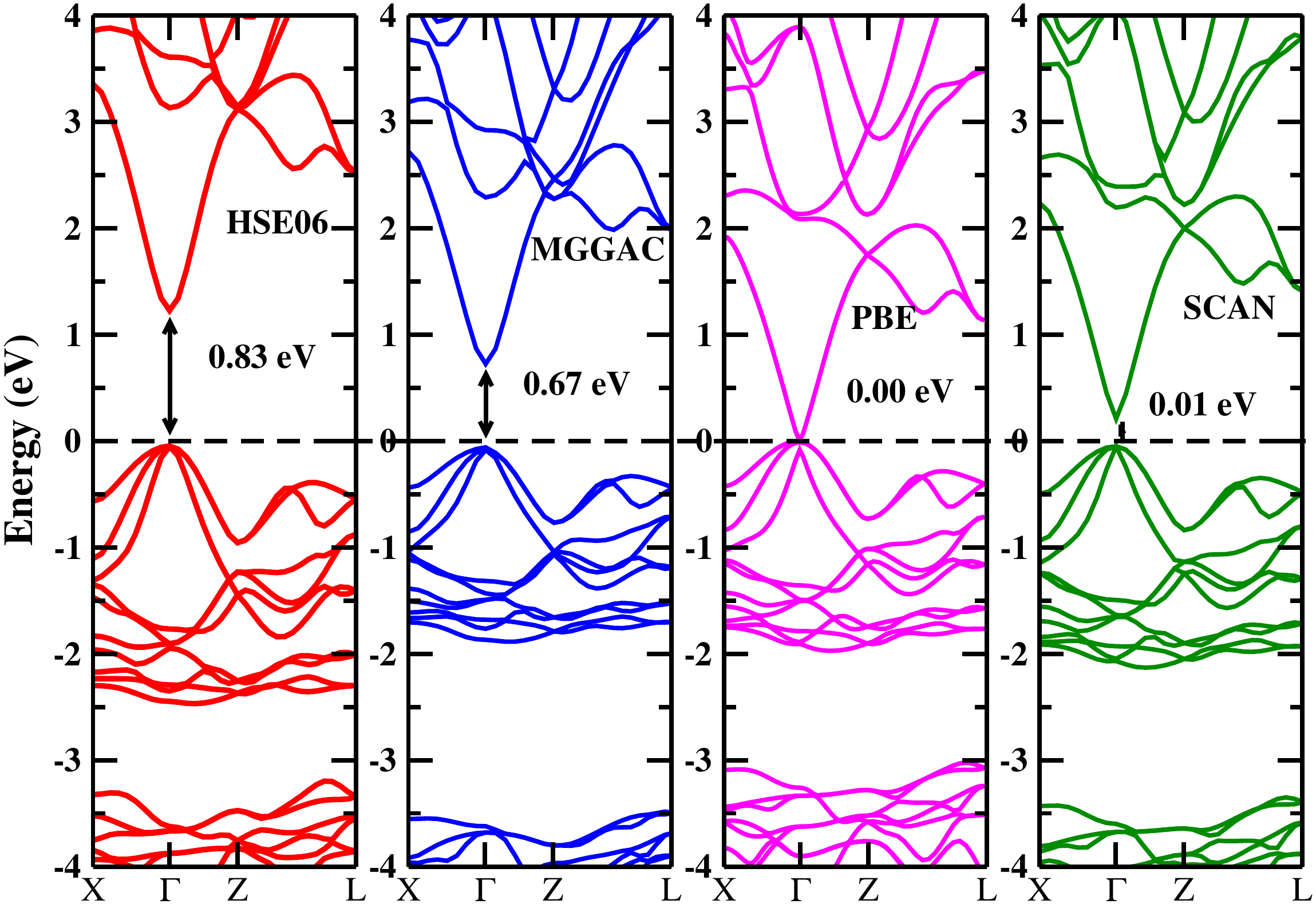}
    \caption{Band structures of CuInS$_2$ as obtained from different methods.
}
  \label{fig-band-1}
\end{figure*}

\begin{figure}
      \includegraphics[width=8.5 cm, height=6.0 cm]{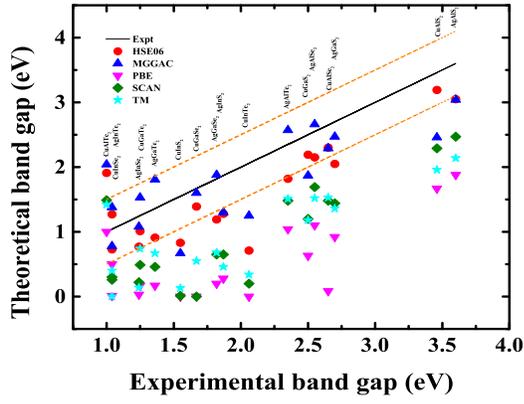}\\
      \vspace{0.0 cm}
      \includegraphics[width=8.5 cm, height=6.0 cm]{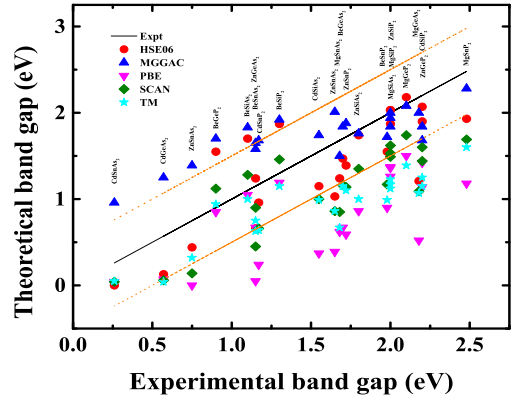}\\
      \caption{Experimental versus theoretical bandgaps of I-III-VI2 (upper panel) and II-IV-V2 (lower panel)
             semiconductors as obtained from different methods. The values within dotted lines indicate
             the bandgaps within the error of $\pm 0.5$ eV of the experimental values. The experimental bandgaps
             of I-III-VI2 are taken from ref.~\cite{xiao2011accurate} and II-IV-V2 from ref.~\cite{shaposhnikov2012abinitio}.}
  \label{fig-bands-all}
\end{figure}

\begin{figure}
  \includegraphics[width=7 cm, height=5 cm]{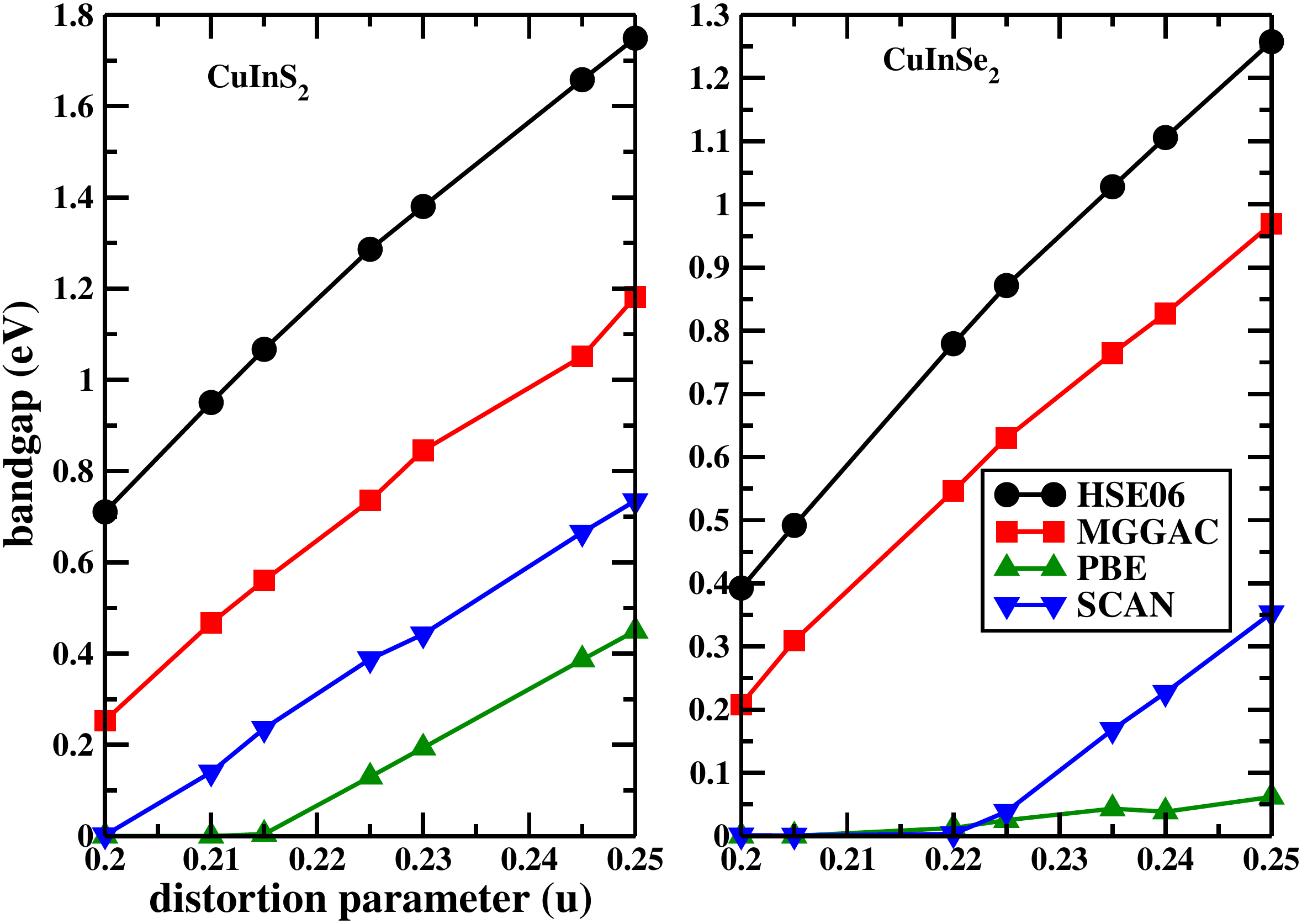}
  \caption{Variation of bandgaps of CuInS$_2$ and CuInSe$_2$ as a function of distortion parameter ($u$).}
  \label{u-vs-band}
\end{figure}

\begin{figure}
  \includegraphics[width=9 cm, height=5 cm]{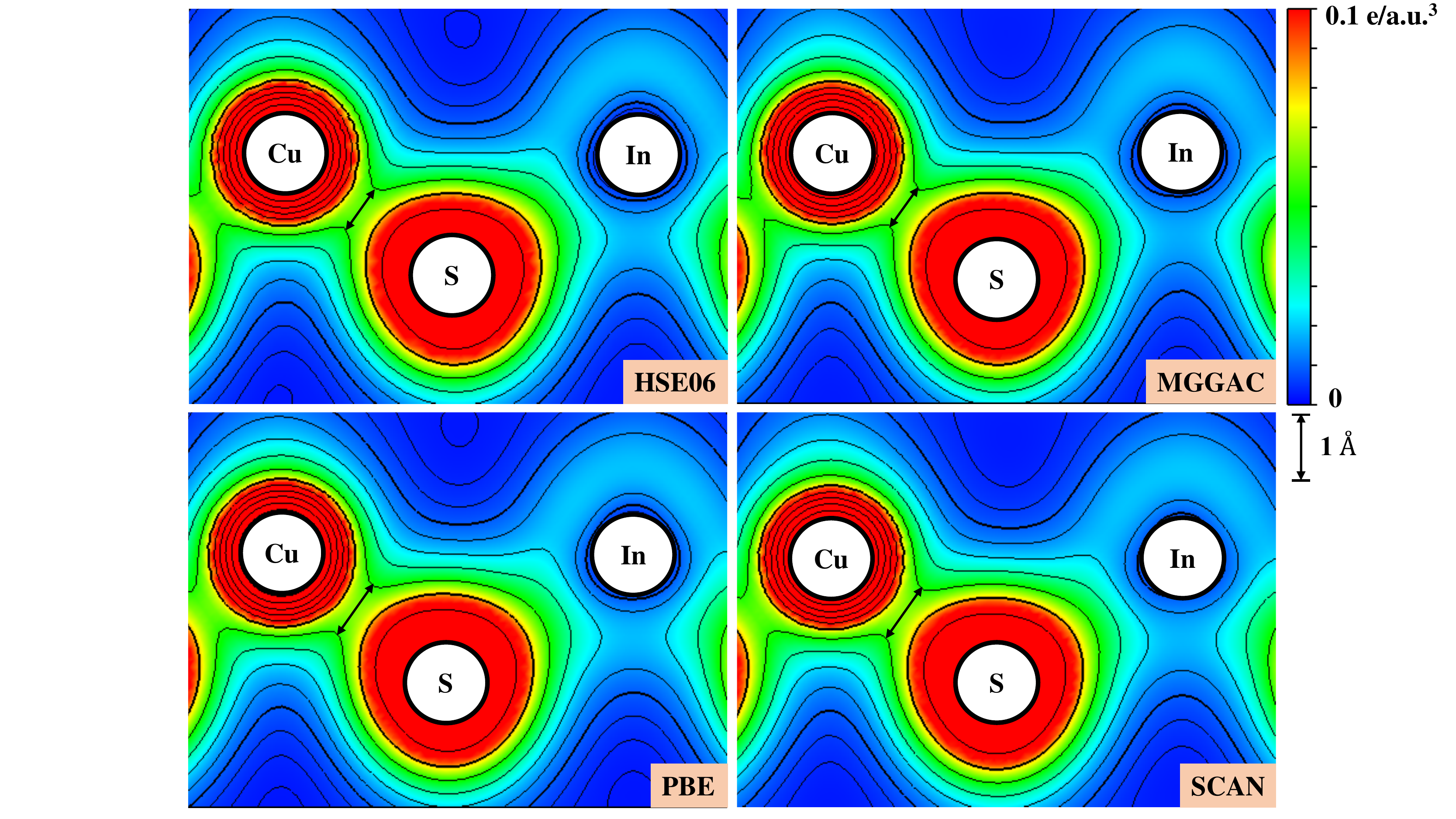}
  \caption{Electronic charge density distribution contours of CuInS$_2$ in the plane contain the Cu, In, and S atoms according to 
  different methods. The logarithmic scale is used for the clear visualization of the isosurfaces.}
  \label{fig-chg-den1a}
\end{figure}

\begin{figure*}
  \includegraphics[width=8.0 cm, height=8 cm]{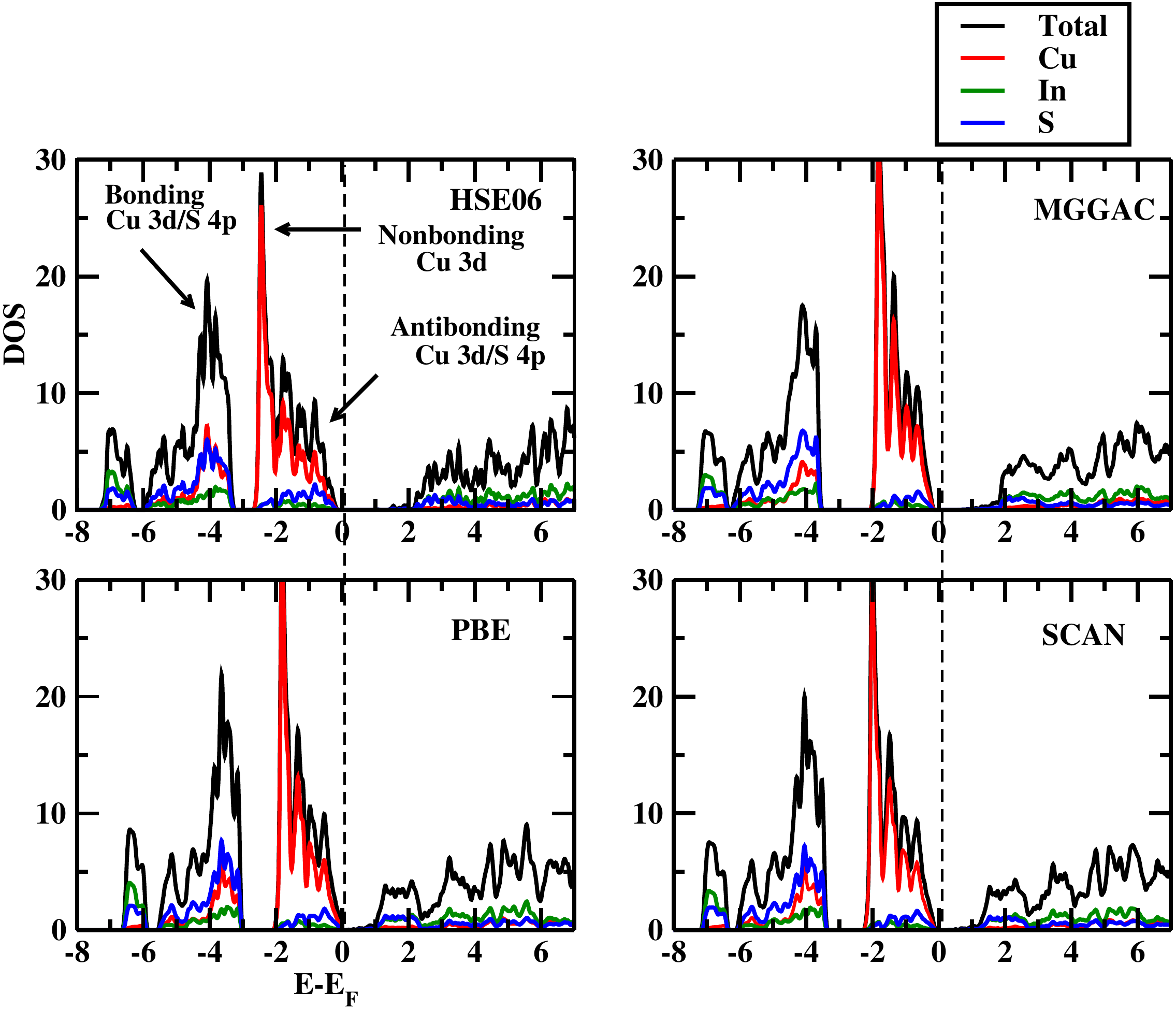}
  \hspace{0.5 cm}
    \includegraphics[width=8.0 cm, height=8 cm]{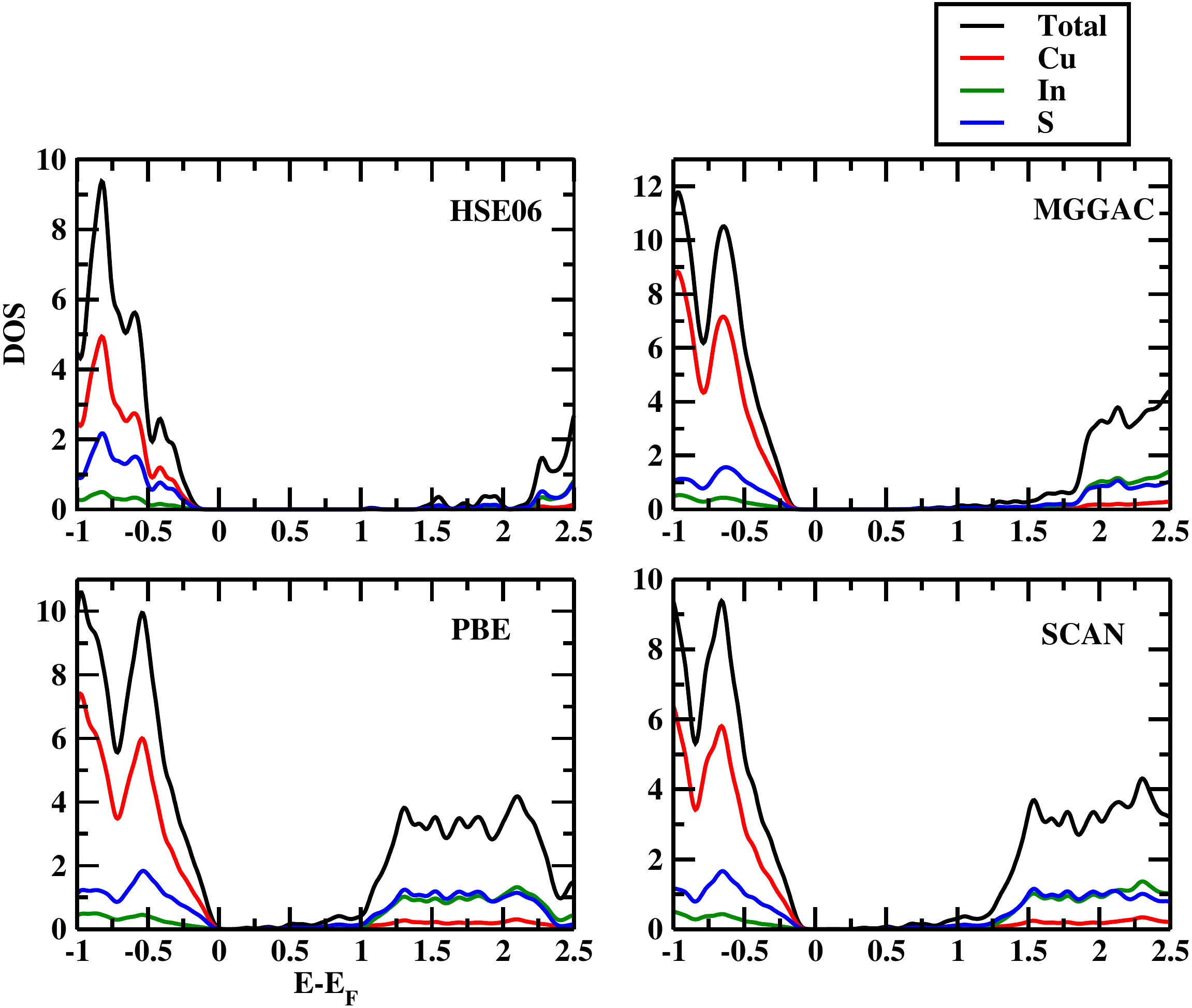}
  \caption{Density of states (DOS) for CuInS$_2$ calculated from different methods.
  Right panel shows the DOS in energy range $-1~eV <(E-E_F)<2.5~eV$.
  The Fermi energy (E$_F$) is set to $0$ eV and is same as E$_{VBM}$.}
  \label{fig-dos-all}
\end{figure*}

The bandgaps of different Cu-based chalcopyrite systems as obtained from different methods
are shown in Table S1 of the supporting information~\cite{support}. As can be seen, the bandgaps computed using hybrid HSE06 
method, are in close agreement with the experimental values. Further, among the semilocal functionals, MGGAC performs 
in a significantly better way than others, as it opens the bandgaps in CuInS$_2$, CuInSe$_2$, and CuInTe$_2$. 
The bandgaps computed using SCAN, TM and PBE are found to be severely underestimated and thus these methods perform poorly. 

Fig~\ref{fig-band-1} shows the band structure of CuInS$_2$ along  high-symmetry directions in BZ. 
As can be seen, the bandgap computed using MGGAC is improved significantly and is quite close to that 
obtained using HSE06. The SCAN method though opens the bandgap for CuInS$_2$, its 
magnitude comes out to be very small. The use of PBE results in zero bandgap.

It may be noted that the HSE06 functional has more often been used to study chalcopyrites  
as compared to other global hybrid functionals such as PBE0, because of the presence of screened Coulomb interaction,
which reduces the computational cost significantly. 

As can be seen in Table S1 of the supporting information~\cite{support},
excellent estimates of electronic properties are obtained from hybrid functional HSE06 
with its default value of $25\%$ fraction of Fock exchange. 

Interestingly, among considered semilocal approximations, the MGGAC functional performs quite closely to HSE06 for most semiconductors. 
(see supplemental material).
A comparative plot of theoretical and experimental bandgaps is shown 
in Fig~\ref{fig-bands-all} which clearly indicates that bandgap estimates obtained using MGGAC are more accurate than 
those obtained using other semilocal functionals.

Based on the comparision of the results presented in this work with those in Ref.~\cite{braga2011panchromatic,shaposhnikov2012abinitio} for I-III-VI2 and II-IV-V2
chalcopyrite semiconductors, the performances of different functionals are twofold:

\begin{itemize}
    \item {\bf I-III-VI$_2$ semiconductors:} The direct bandgaps are found to be most severely underestimated when computed using semilocal 
    functionals like GGA, SCAN, and TM. 
    For Ag based chalcopyrites too, a similar trend is observed.     
   A consistent improvement in the bandgap magnitudes is observed, 
   when computed using the MGGAC functional. This is expected due to the 
   improvement in the description of $p-d$ hybridization discussed in the previous section. In the case of CuAlTe$_2$, 
    the bandgap computed using MGGAC and HSE06 is found to be overestimated whereas PBE-GGA 
    estimate of the bandgap is closer to the experiment. Further, for all the Te based systems, MGGAC overestimates the bandgaps as 
    compared to HSE06. In particular, the MGGAC overestimates the bandgaps of CuAlTe$_2$ and AgInTe$_2$ by a large margin. 
    Nevertheless, for other Te based systems, the performance of MGGAC is reasonably good with bandgap estimates closer  
    to the experiments as compared to other semilocal methods.

    \item {\bf  II-IV-V2 semiconductors:} 
      The computed GGA bandgaps are found to be in excellent agreement with
      the experiments~\cite{shaposhnikov2012abinitio} and earlier reported theoretical results~\cite{shaposhnikov2012abinitio}. 
     The MGGAC computed bandgaps are found to be comparable with HSE06 computed bandgaps. 
    However, both MGGAC and HSE06 overestimate the bandgaps in case of BeSiAs$_2$, BeSiP$_2$, 
    CdGeAs$_2$, CdSnP$_2$. In case of these semiconductors, the SCAN functional performs well 
    as compared to other semilocal (MGGAC, TM, GGA) methods (see ref.~\cite{support} for details). 
    As can be seen in Fig~\ref{fig-bands-all}, the MGGAC computed bandgaps are overestimated by more than 0.5 eV in case of
    CdSnAs$_2$, CdGeAs$_2$, ZnSnAs$_2$, BeGeP$_2$, BeSiAs$_2$, 
    and BeSiP$_2$.    
    
\end{itemize}

It may be noted that for chalcopyrite semiconductors containing Be and Cd, the HSE06 bandgaps are comparable to the GW$_0$ bandgaps 
as reported in Ref.~\cite{shaposhnikov2012abinitio}. Further, the semilocal functionals SCAN, TM, GGA underestimate the bandgaps. 
However, MGGAC overestimates the bandgap for BeSiAs$_2$, BeSiP$_2$, CdGeAs$_2$, CdSnP$_2$. In case of BeGeAs$_2$ and BeSnP$_2$, MGGAC 
bandgaps are comparable to the experimental values and are in better agreement than that for HSE06 functional.

For semiconductors containing Mg and Zn, all functionals estimate bandgaps close to the experimental values except for some semiconductors
containing Zn.

Overall, it may be concluded from the calculations and above discussions that the bandgaps estimated by semilocal MGGAC are comparable to 
those estimated by hybrid functional HSE06 to a great extent with a minimal deviation (see Fig~\ref{fig-bands-all}).

To visualize the variation of the bandgap with the distortion parameter ($u$), we plot in Fig.~\ref{u-vs-band} the bandgaps of 
CuInS$_2$ and CuInSe$_2$ computed using different functionals as a function of $u$. As can be seen, the bandgap increases almost 
linearly with $u$ in range 0.2 $<$ $u$ $<$ 0.25. For both the chalcopyrite systems, HSE06 bandgaps are improved over MGGAC
bandgaps, which in turn are improved over SCAN bandgaps. For CuInS$_2$ and CuInSe$_2$, the SCAN bandgap opens for $u>0.2$ and
$u>0.22$, respectively. As can be seen, the MGGAC and HSE06 bandgaps are finite for all values of $u$ (in range shown)
for both the chalcopyrites. The difference in the bandgaps obtained from different methods is expected due to strong interplay between 
the structure parameter $u$ and the hybridization of the $p-d$ orbitals~\cite{vidal2010strong}. We will revisit
this point later in this paper.

In supporting information~\cite{support}, we also discuss and analysis another chalcopyrite system CuInSe$_2$. As can be seen in supporting information~\cite{support}, the 
MGGAC and HSE06 open the bandgap of this system, whereas it remains closed for PBE and SCAN functionals.

\subsection{ Charge densities}

The charge density contours for CuInS$_2$ are shown in Fig.\ref{fig-chg-den1a}, which reveals the the presence of both ionic and covalent bonds. 
The mixed bonding is seen irrespective of the XC functionals used. As can be seen, the electron density is more enhanced 
towards the side of the S-atom than that of the In-atom, which indicates that the In$-$S bond is more ionic.
Further, the even distribution of the electron density in the case of the Cu$-$S bond indicates that the bonding is more covalent. The arrow labeled in the Cu$-$S bond (see Fig. \ref{fig-chg-den1a}) indicates the isosurface value.  
The size of the arrow is larger in the case of PBE and SCAN than the one for MGGAC and HSE06, which follows the bandgap enlargement rule as well as the differences in the $u$ parameter as the arrow size is related to the corresponding bond length and varying degrees of $p-d$ hybridization for these chalcopyrites. From Table S1 of the supporting information~\cite{support}, 
it can be seen that the calculated $u$ parameter, which depends on the bond length of Cu$-$S, is less for PBE and SCAN (0.218) than that for MGGAC (0.224) and HSE (0.226). The greater the isosurface arrow, the lower the atomic distance is, 
which indicates greater $p-d$ repulsion and thereby smaller bandgap. It is generally understood that methods (within the DFT framework) that do not take into account the self-energy correction, underestimate the bandgaps. However, in the case of MGGAC, the presence of internal screened coulomb energy $U$ results in bandgap estimates which are comparable to bandgaps obtained 
using hybrid functionals as well as state-of-the-art many-body perturbation technique. We will revisit and elaborate more on 
this point in a later section. The MGGAC computed bandgaps also exhibit general trend which is similar to that seen for 
bandgaps obtained using other functionals. For instance, the computed bandgaps of semiconductors containing $P$-atoms are 
always larger than bandgaps of semiconductors containing As-atoms.

\subsection{ Density of states}

The electronic structures of all chalcopyrite semiconductors are qualitatively similar. In the following, we present and discuss the electronic structure of CuInS$_2$. The electronic structure of CuInSe$_2$ is presented in the supporting information~\cite{support}. The total density of states (DOS) and partial DOS as computed with different XC functionals are 
shown in Fig~\ref{fig-dos-all} and Fig. S3 of supporting information~\cite{support} for CuInS$_2$ 
and CuInSe$_2$, respectively. As can be seen, the contribution from the valence Cu-$4s$ and Cu-$3p$ states near the VBM is negligible as compared to Cu-$3d$ states. The valence band region exhibits two main groups: up to around $-2$ eV from the VBM,
the region is mainly composed of $p$ and $d$ states with primary contribution from Cu$-3d$ states. The deeper valence band in 
the energy range -$2$ eV to $0$ eV results from strong hybridization of Cu$-3d$ and S$-3p$ states. The conduction band region is
contributed by In$-5p$ and S$-3p$ states along with a minor contribution of Cu$-3d$ states. The dispersion of the valence and
conduction bands are found to be almost the same for all semilocal functionals (GGA, SCAN, and MGGAC)
with an exception that the band width (energy between the highest and lowest allowed levels) for MGGAC is
wider than that for PBE and SCAN.

To encapsulate the behavior of the VBM and CBM, we show the DOS plot for the respective functionals in the right panel of  Fig~\ref{fig-dos-all},
in the energy range $-1~eV <(E-E_F)<2.5~eV$. The enhancement in the bandwidth is evident in case of MGGAC as compared to that for SCAN and PBE 
semilocal functionals. In case of MGGAC, the CBM shifts to the higher energy level due to slightly enhanced hybridization between Cu$-3d$ and S$-3p$ 
states at VBM, resulting in the opening of the bandgap. However, in case of PBE and SCAN,
the $3d$ non-bonding region lies a little bit of the higher side on the energy scale, which results in the reduction of the bandgap
as compared to that in case of MGGAC and HSE06.

In the case of HSE06, the relative bandgap enhancement compared to that for semilocal functionals happens because of the inclusion of the non-local 
HF exchange, which reduces the repulsion between Cu$-3d$ and S$-4p$ states and thereby results in the bandgap opening~\cite{kim2016screened}. 
Importantly, the bandgap enlargement in the case of HSE06 as compared to other semilocal functionals (MGGAC, SCAN, TM, and PBE) may be attributed to
($i$) enlargement of bandwidth and the orbital spacing which directly influences the bandgap and ($ii$) the shift of non-bonding $3d$ 
states downward which reduce the $p-d$ repulsion. 


Nevertheless, the performance of the MGGAC within semilocal XC functionals is quite encouraging and the same trend of results 
also follows for all I-III-VI2 semiconductors as the valence band maximum is influenced by the $4d$ states of I-group atoms.  
On the other hand, in the case of II-IV-V2 semiconductors, the bandgap character is not influenced by $d$ states of II-group atoms, as
these are highly localized and contribute to valence bands far from VBM~\cite{shaposhnikov2012abinitio}. 

In supporting information~\cite{support}, we show the DOS and PDOS for CuInSe$_2$. In this case also, the opening of the bandgap can be explained 
sililarly as discussed for CuInS$_2$. The reduction in repulsion between Cu$-3d$ and Se$-4p$ states is responsible for the opening of bandgap. 
For MGGAC, this repulsion reduces at the VBM and the CBM is shifted towards the higher energy. This results in the opening of the bandgap more than 
that in case of SCAN.

\section{Formation enthalpies}

\begin{table*}
\begin{center}
\caption{\label{tab-enthalpy} Enthalpy formation energies (eV/f.u.) of CuInSe$_2$ and AgInSe$_2$ as obtained 
from different methods. PBE and HSE06 values are taken from ref.~\cite{kim2016screened}. Numbers in the parentheses
represent the percentage deviation.}
\begin{ruledtabular}
\begin{tabular}{lcccccccccccccccccccccccc}
 Solids&PBE&SCAN&TM&MGGAC&HSE06&Expt.~\cite{Berger_book}  \\
\hline
CuInSe$_2$&-1.78&-2.02&-1.39&-2.35&-2.40&-2.77\\
          &[-35.6]&[-27.1]&[-49.5]&[-15.1]&[-13.4]&\\
AgInSe$_2$&-1.71&-1.90&-1.26&-2.25&-2.22&--2.51\\
          &[-31.8]&[-24.3]&[-49.5]&[-10.3]&[-11.4]&\\
\end{tabular}
\label{enthalpy}
\end{ruledtabular}
\end{center}
\end{table*}

\begin{figure}
  \includegraphics[width=7 cm, height=6 cm]{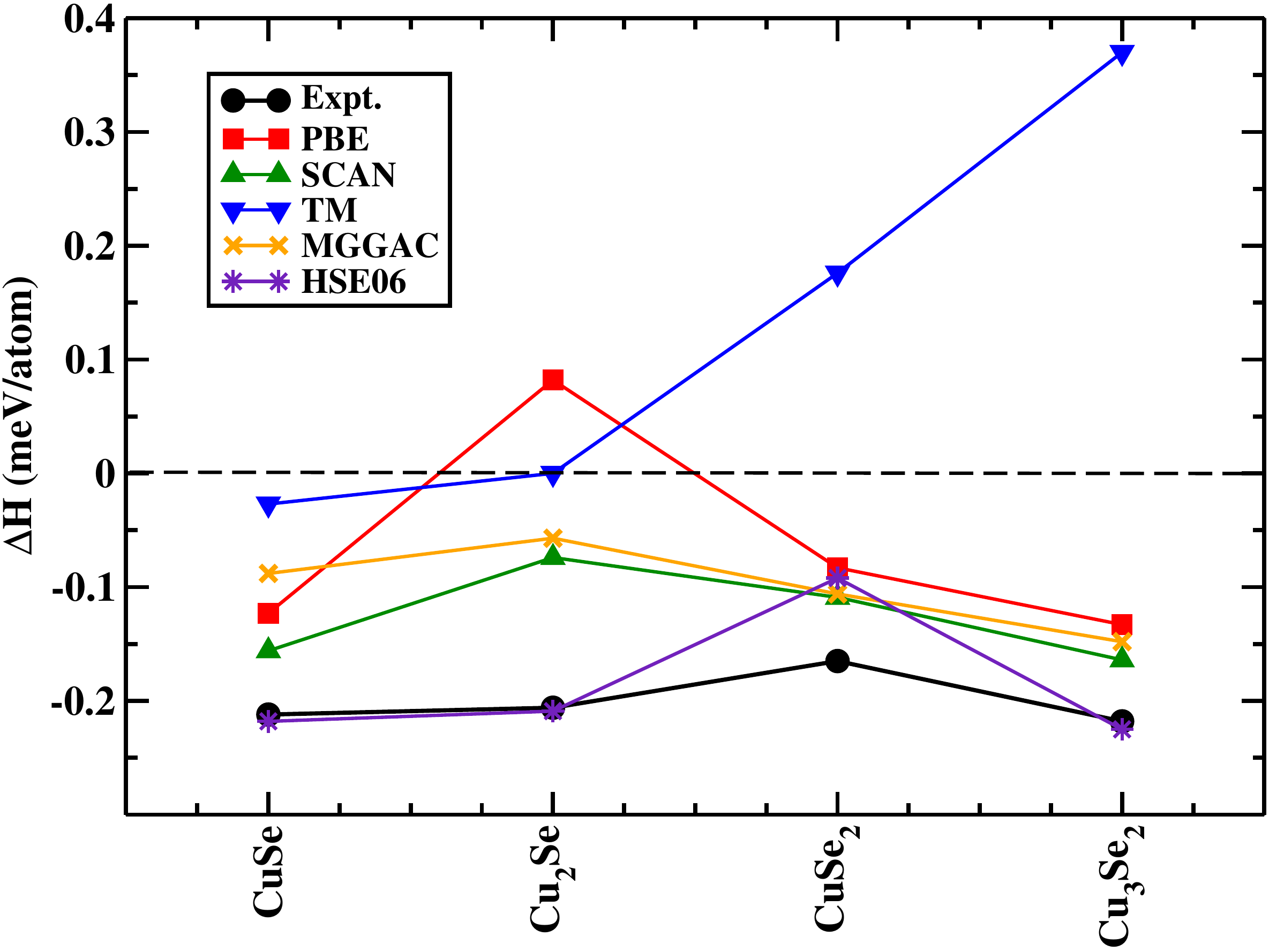}
  \caption{Enthalpy formation energies (meV/atom) of Cu-Se binaries as obtained 
  from different methods}
  \label{figenthpy}
\end{figure}

We now discuss the enthalpy formation energies of chalcopyrites and competing binary phases of Cu-Se, computed using different meta-GGA functionals. In the thermodynamical framework, the enthalpy formalism energy per atom for a chalcopyrite system
is defined as:
\begin{equation}
 \Delta H= [\epsilon_{system} - \sum_{i}n_i\epsilon_i]/m~,
\end{equation}
where $\epsilon_{system}$ is the energy of the chalcopyrite system, $n_i$ and $\epsilon_i$ are the number and energy of the $i^{th}$ constituent in the system, and $m$ is the total number of atoms in the system. 
All energies are calculated in their stable solid form. A negative (positive) enthalpy formation energy 
indicates that the system is stable (unstable).

The Table~\ref{enthalpy} shows the formation enthalpies of CuInSe$_2$ and AgInSe$_2$ computed using different 
functionals. For these solids, the experimental formation enthalpies have been reported
in ref.~\cite{Berger_book}. The PBE and HSE06 results in Table~\ref{tab-enthalpy} are taken from ref.~\cite{kim2016screened}. 
As can be seen, the PBE underestimates the formation energies of both the chalcopyrites by more than $30\%$. 
The SCAN method improves the enthalpy estimates as compared to those obtained using PBE. 
The SCAN values are underestimated by less than $30\%$. Further, Table~\ref{tab-enthalpy} shows that the formation enthalpy estimates are improved significantly with the use of MGGAC functional, which performs as accurately as hybrid HSE06 functional and with computational cost much lesser than that of HSE06.

It may be noted that the PBE functional does not properly describe different bondings such as metallic, covalent, and/or ionic 
in different solids~\cite{stevanovimmode2012correcting,jain2011formation,hautier2012accuracy}. The bonding nature is much better
described by meta-GGA\cite{sun2013density} functionals, which is reflected in the performance of SCAN and MGGAC. 
This is because the meta-GGA functionals are generally better than GGAs in describing the covalent, metallic, and 
non-covalent interactions ~\cite{sun2013density,della2016kinetic}. 

Incidentally, the TM functional fails drastically in providing reasonable estimates of formation energies as it exhibits error more than PBE. This failure of the TM may be attributed to its order-of-limit problems, which is important for
estimating the formation of enthalpies~\cite{patra2020way}. In case of the HSE06, the good performance is due to 
reasonably correct desprition of the $d$-states of Cu and Ag atoms.

The formation enthalpy of CuInSe$_2$ computed using MGGAC and HSE06 is in close agreement with the experiments, 
which is expected due to improved estimates of energy eigenvalues, In-4$d$ states, band structure and DOS in it 
(see supporting material~\cite{support}).

Next, we assess the functionals performance for the several competing candidate binary phases of Cu-Se. 
The formation enthalpies per atom for different Cu$-$Se phases are plotted in Fig.~\ref{figenthpy}.
As expected, PBE underbinds the formation enthalpies, with a positive $\Delta H$ for Cu$_2$Se compound. 
The HSE06 estimated formation energies are found to be in closest agreement with the experimental values. 
The SCAN and MGGAC perform closely, albeit SCAN performance slightly better than MGGAC in all cases. 
This may be partially attributed to the metallic character of the CuSe, CuSe$_2$, and Cu$_3$Se$_2$, 
which is better described by the SCAN functional as it takes into account the slowly varying second/fourth order 
terms in the gradient expansion of the exchange.  
Furthermore, the SCAN correlation is free from spurious one-electron self-interaction which tends to improve the
formation enthalpy estimates. Nevertheless, both SCAN and MGGAC formation energies are negative (including 
semiconductor Cu$_2$Se) and therefore are much better estimated as compared to those by PBE-GGA. 
Fig.~\ref{figenthpy} also shows that the performance of the TM functional is quite poor for all the binary 
phases of Cu-Se. As can be seen, the TM functional incorrectly predicts CuSe$_2$ and Cu$_3$Se$_2$ as unstable. 
This significantly poor performance of the TM functional in providing formation enthalpy estimates
may be attributed to its order-of-limit problem~\cite{patra2020way}.

\section{Relation between performance and mathematical features of
functionals} 

\begin{figure}
  \includegraphics[width=8 cm, height=6 cm]{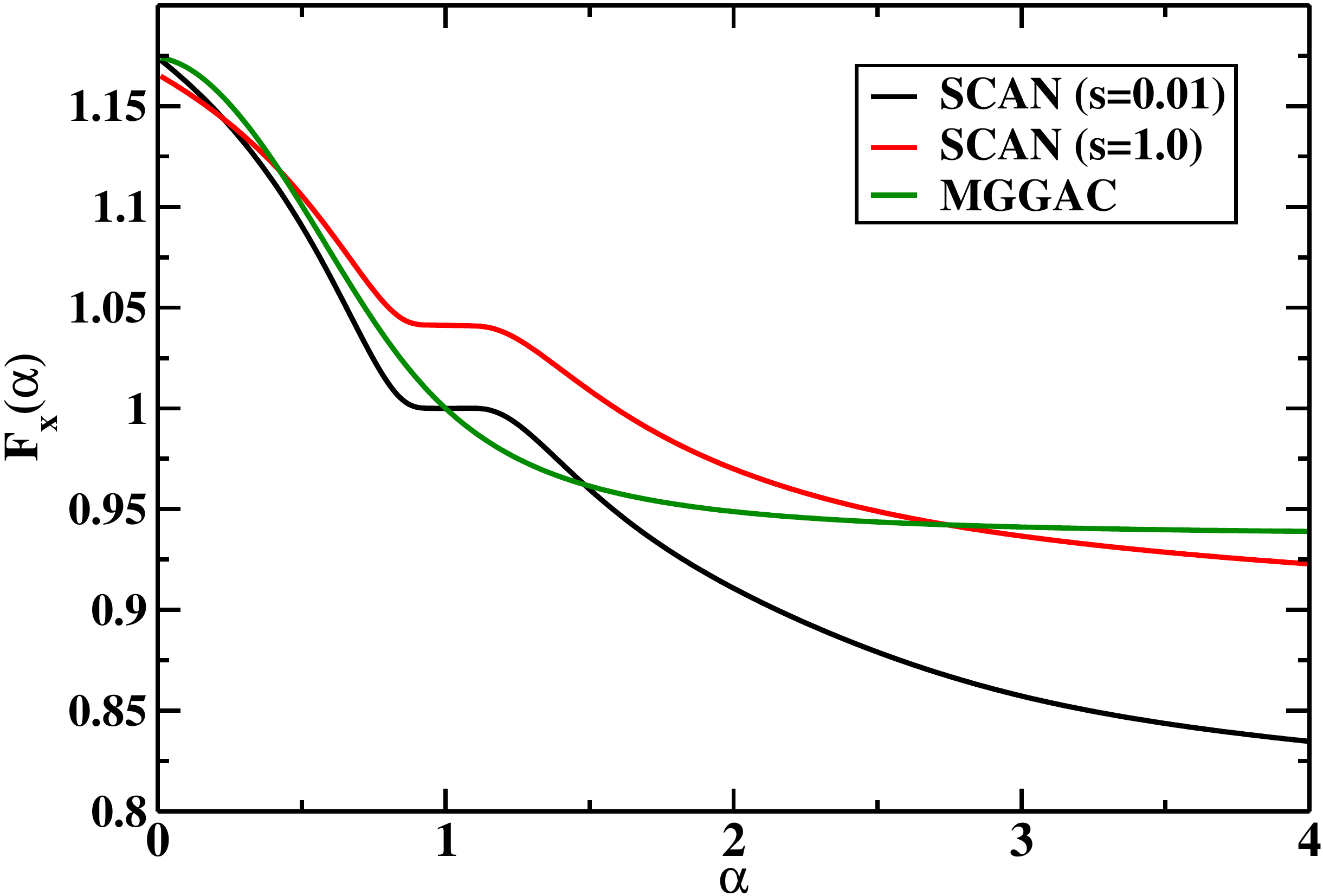}
  \caption{The exchange enhancement factor of SCAN and MGGAC as a function of 
  $\alpha$ for different values of reduced density gradient $s$. The MGGAC exchange is 
  independent of $s$.}
  \label{fig_enh}
\end{figure}


Next, we discuss the plausible reasons behind improvement in performance of MGGAC functional 
over GGA or SCAN  methods. 
To analyze the results, let us consider the behavior of the exchange enhancement factor ($F_x$) of SCAN and 
MGGAC functionals, the gKS potential of the meta-GGA functionals, and its relation to the derivative discontinuity 
($\Delta_x$), which is connected with the localized $d$- or $f$ -states, often encountered
in the chalcopyrite systems. 
In Fig.~\ref{fig_enh}, we plot the $F_x$ as a function of meta-GGA ingredient $\alpha$
for different values of reduced density gradient $s$. 
The MGGAC $F_x$ does not depend on $s$ and is much smoother 
than that for 
SCAN and its slope $\partial F_x/\partial\alpha$ is more negative in the energetically relevant regions
of solids ($0<s \leq 1$) resulting in inclusion of more $\Delta_x(\propto \partial F_x/\partial\alpha$). 
In fact the 
negative slope $\partial F_x/\partial\alpha < 0$ is also connected to the ultranonlocality effect which results in
polarization 
effect in solids~\cite{aschebrock2019ultranonlocality}. 
Furthermore, the $\alpha$ is also interpreted as the measure of the 
well-known electron localization factor via the exchange energy, similar to the EXX~\cite{aschebrock2019ultranonlocality}. 
Importantly,
because of the more enhanced $\alpha$ dependent and $\partial F_x/\partial\alpha < 0$, 
some amount of inherent on-site Coulomb interaction ($U$) is included in
the orbital energies and eigenvalues
corresponding to MGGAC functional.
The on-site Coulomb interaction ($U$) is 
important to treat the 
highly delocalized band states of rare-earth elements and strongly localized late transition metal elements.
It may be noted 
that the SCAN functional also includes some amount of inherent $U$~\cite{peng2017synergy} 
which results in improved performances 
as compared to PBE+U
and SCAN-no-U~\cite{sai2018evaluation,varignon2019mott,kirchnerhall2021extensive,wexler2020exchange,long2020evaluating}. 
However, because of the more 
realistic and enhanced $\partial F_x/\partial\alpha < 0$, the MGGAC is a slightly better in treating the $d$- and $f$- orbitals and its results 
for chalcopyrite systems 
are
often close to the 
HSE06, where the $25\%$ HF exact exchange is used to treat the strongly localized $d$- bands. 
It may be noted
that along with the 
localized $d$- states, the $s$- and $p$-states are also well described by the MGGAC.
From the structure of the DOS plot (see Fig~\ref{fig-dos-all} and Fig. S3 of supporting information~\cite{support}) obtained using 
the MGGAC functional, one can readily observe that it treats the non-bonded 
Cu$-3d$ and hybridized $3d-4p$ (of Cu and Se) 
states in a bit better way than PBE and SCAN
functionals. This results in the reduction of strong localization 
effect of $3d$ states, which in turn opens a sizable bandgap between occupied and unoccupied sub-bands. 

In general,
the success and  overall improvement of 
MGGAC functional vis-a-vis
LDA, GGA, and SCAN functioanls are retained
for other chalcopyrite systems discussed in this article.

\section{Concluding remarks}

The comparative assessment of the structural properties,  electronic properties, and formation energies of Cu-based multinary semiconductors is performed using GGA, meta-GGA, and hybrid density functionals within the framework of density functional theory. In particular, we assess the performances of SCAN, TM, and MGGAC meta-GGA functionals, which are very recent 
and reasonably successful exchange-correlation (XC) functionals in predicting various solid-state properties. 
To assess the structural properties, the tetragonal distortion ($\eta$) and/or anion displacement ($u$) 
are first computed from different functionals. 
The PBE-GGA functional is found to be inadequate to estimate accurately the electronic structures and associated 
properties of Cu-based chalcopyrite systems due to its inability to describe the intrinsic localization and/or strong 
correlation of the $d$ electrons.

Our results for PBE and HSE06 are in agreement with earlier reported studies, wherein it was shown that the HSE06 generally
improves over PBE because of less delocalization error, which in turn improves the bandgap, and density of state (DOS) 
estimates. 

The estimates of various quantities computed using MGGAC and SCAN are found to be significantly improved as compared to those 
obtained using GGA. Further, the performances of MGGAC is found to be encouraging among meta-GGA functionals and 
quite close to that of the HSE06 in providing the estimates of $\eta$, $u$, band structures, and enthalpy formation energies
of Cu-based multinary semiconductors. 
The reasonably improved and better performance of MGGAC is attributed to its
a better description of strongly localized $d$ electrons in these systems.

In addition, the performance of the aforementioned functionals is also explored
for several competing candidate binary phases of Cu-Se.
For these phases also, the MGGAC is found to be quite accurate.
However, the SCAN functional performs slightly better than MGGAC for binary phases.
The performance of TM functional is found to be poor in comparison to other meta-GGA functionals possibly due to
its order of limit problem.
Overall, our results strongly suggest that MGGAC (meta-GGA) functional can be highly useful for studying 
Cu-based multinary semiconductors as well as other systems with 
$d$-electrons, due to its good performance and low computational cost.

\section{Acknowledgemets}
A.G. would like to thank INSPIRE fellowship, DST, India for financial support.
S.J. would like to thank NISER, Bhubaneswar for partial financial support. Calculations are performed high performance computing
(HPC) clusters of IITH, Hyderabad. Part of calculations are also performed in
the KALINGA and NISERDFT high performance computing
(HPC) clusters of NISER, Bhubaneswar.


\twocolumngrid
\bibliography{reference.bib}
\bibliographystyle{apsrev.bst}

\end{document}